\documentclass[12pt,preprint]{aastex}
\usepackage{epsfig}
\shorttitle{Be Star HD 215227}
\shortauthors{Williams et al.}

\begin{document}


\title{The Be Star HD 215227: A Candidate Gamma-ray Binary} 

\author{S. J. Williams\altaffilmark{1}, D. R. Gies, 
  R. A. Matson, Y. Touhami}
\affil{Center for High Angular Resolution Astronomy and
 Department of Physics and Astronomy,
 Georgia State University, P. O. Box 4106, Atlanta, GA 30302-4106; 
 swilliams@chara.gsu.edu, gies@chara.gsu.edu, rmatson@chara.gsu.edu, 
 yamina@chara.gsu.edu}

\altaffiltext{1}{Guest investigator, Dominion Astrophysical Observatory, 
Herzberg Institute of Astrophysics, National Research Council of Canada.}

\author{E. D. Grundstrom}
\affil{Physics and Astronomy Department, Vanderbilt University, 
6301 Stevens Center, Nashville, TN 37235; erika.grundstrom@vanderbilt.edu}

\author{W. Huang}
\affil{Department of Astronomy, University of Washington, 
Box 351580, Seattle, WA 98195-1580; hwenjin@astro.washington.edu}

\author{M. V. McSwain}
\affil{Department of Physics, Lehigh University, 16 Memorial Drive E., 
Bethlehem, PA 18015; mcswain@lehigh.edu}

\slugcomment{V2. 09/03/2010 -- Submitted to ApJL}


\begin{abstract}
The emission-line Be star HD~215227 lies within the positional 
error circle of the newly identified gamma-ray source AGL~J2241+4454.
We present new blue spectra of the star, and we point out the 
morphological and variability similarities to other Be binaries. 
An analysis of the available optical photometry indicates a
variation with a period of $60.37 \pm 0.04$~d, which may 
correspond to an orbital modulation of the flux from the disk 
surrounding the Be star.  The distance to the star of 2.6~kpc 
and its relatively large Galactic latitude suggest that the 
binary was ejected from the plane by a supernova explosion 
that created the neutron star or black hole companion. 
The binary and runaway properties of HD~215227 make it 
an attractive candidate as the optical counterpart of 
AGL~J2241+4454 and as a new member of the small class of 
gamma-ray emitting binaries. 
\end{abstract}


\setcounter{footnote}{1}

\section{Introduction}                              

Gamma-ray binaries are a class of high energy and very-high energy 
emitting sources that consist of a massive star and compact companion
(Mirabel 2007; Dubus et al.\ 2010; McSwain 2010).  Six such objects 
are known sources of TeV emission: LS~5039, Cygnus~X-1, 
Cygnus X-3, LS~I~+61~303, PSR~B1259$-$63, and HESS~J0632+057.  
The massive star component is a luminous O-star in the first two, 
a probable WR star in the third, and a Be star in the last three cases.
All these massive stars have winds and the Be stars also eject mass into 
an outflowing circumstellar disk.  The interaction of this mass loss 
with a degenerate companion can lead to gamma-ray emission in several 
ways (Parades 2008).  First, if the companion is a pulsar, then a high speed 
wind from the mass donor can collide with the pulsar wind in a shock region, 
and inverse Compton scattering of stellar photons with relativistic 
electrons in the shock can create gamma-rays (Dubus et al.\ 2010). 
Second, if the companion is a black hole, then gas accretion can lead to 
the formation of relativistic jets, and gamma-ray emission may
occur by inverse Compton scattering from jet electrons and/or by the 
decay of neutral pions that originate in inelastic proton - proton 
collisions (Romero et al.\ 2007).  

Recently, Lucarelli et al.\ (2010) announced the detection with 
the AGILE satellite (Tavani et al.\ 2009) of gamma-ray emission above 100~MeV 
from a new unidentified source, AGL~J2241+4454.  The source has 
Galactic coordinates of $(l,b)=(100\fdg0,-12\fdg2)$ with an error circle 
radius of approximately $0\fdg6$.  
The source has not yet been detected by the NASA Fermi Gamma-Ray 
Observatory\footnote{Fermi LAT Report, 2010 July 30; http://fermisky.blogspot.com/}.
The AGILE point sources found to date 
include pulsars, blazars, supernova remnants, and high mass X-ray binaries 
(Pittori et al.\ 2009), but there are no cataloged examples of any of
these in the region close to AGL~J2241+4454 (Lucarelli et al.\ 2010). 
We point out two possibilities that deserve further attention.  
First, Brinkmann et al.\ (1997) found an X-ray source and probable quasar 
in this region, RXJ$2243.1+4441$, that may be a possible source of gamma-ray 
flaring (Vercellone et al.\ 2008).  However, there is no known optical 
counterpart (radio designation B$3~2241+444$), which would
probably appear at $V \approx 16$ (based upon the observed X-ray flux and an 
assumed frequency power law with $\alpha = 1.3$).  Further multiwavelength 
observations are required to say anything more about the candidacy of this object. 
Second, the area contains the Galactic Be star HD~215227 (BD~+43$^\circ$4279, 
HIP~112148, MWC~656), which is located at $(l,b)=(100\fdg1755,-12\fdg3985)$. 
This object was discovered as an emission-line star by Merrill \& Burwell (1943)
who estimated a spectral type of B0e, and the star was subsequently assigned types
B5:ne (Harris 1955), B3ne$\gamma$ (Petrie \& Lee 1965), and back to B0 
(Hern\'{a}ndez et al.\ 2005).  The classification is difficult because the
spectral lines are very broad and weak (the projected rotational velocity is  
$V\sin i = 262\pm 26$ km~s$^{-1}$; Yudin 2001) and often blended with emission features.  
Here we present new blue spectra of the target, and we argue that HD~215227
is a potential optical counterpart of AGL~J2241+4454 based upon its probable 
runaway and binary character.  


\section{Spectroscopic Observations}         

We obtained blue spectra of HD~215227 with the HIA 
Dominion Astrophysical Observatory 1.8~m telescope on 2010 July 28 and 29.
These observations were made with the Cassegrain 
spectrograph\footnote{https://www.astrosci.ca/DAO/dao72.html} 
with grating 1200B (1200 grooves mm$^{-1}$) in first order, 
and the spectra cover the range 4260 -- 4669 \AA .
The detector was the SITe-2 CCD 
(a $1752\times 532$ pixel array with $15\times 15$ $\mu$m pixels), 
and the resulting spectra have a resolving power of 
$R = \lambda / \Delta \lambda = 4290$ as measured from the
Fe~Ar comparison lines.  Exposures were 300~s in duration, 
leading to spectra with a S/N = 100 per pixel in the continuum.  
The spectra were extracted and calibrated using
standard routines in IRAF\footnote{IRAF is distributed by the National
Optical Astronomy Observatory, which is operated by the Association of
Universities for Research in Astronomy, Inc., under cooperative agreement
with the National Science Foundation.}, and then each 
continuum-rectified spectrum was transformed to a common 
heliocentric wavelength grid in $\log \lambda$ increments. 

The two spectra are illustrated in Figure~1.  There is clear evidence of 
double-peaked emission from the circumstellar disk that is seen in the
core of H$\gamma$ and in numerous \ion{Fe}{2} emission lines.  
The only clear photospheric lines are those of \ion{C}{2} $\lambda 4267$
and \ion{He}{1} $\lambda\lambda 4387, 4471$.  We searched through the 
Be star spectral survey of Grundstrom (2007), and the closest matching 
spectrum we found is that of HR~2142 = HD~41335, which is shown for comparison 
in the top section of Figure~1.  HR~2142 is a well known Be binary system 
that shows deep, narrow, ``shell'' absorption features at certain 
phases in its 81~d orbit (Peters 1983).  The most prominent shell 
feature appearing in Figure~1 is the sharp absorption core of 
H$\gamma$ $\lambda 4340$.  Such absorptions form in cooler, dense 
disk gas seen in projection against the stellar photosphere (Hanuschik 1995),
and their strength may be modulated by azimuthal asymmetries in disk 
density that can occur in Be binaries (Oktariani \& Okazaki 2009). 

\begin{figure}
\begin{center}
{\includegraphics[angle=90,height=12cm]{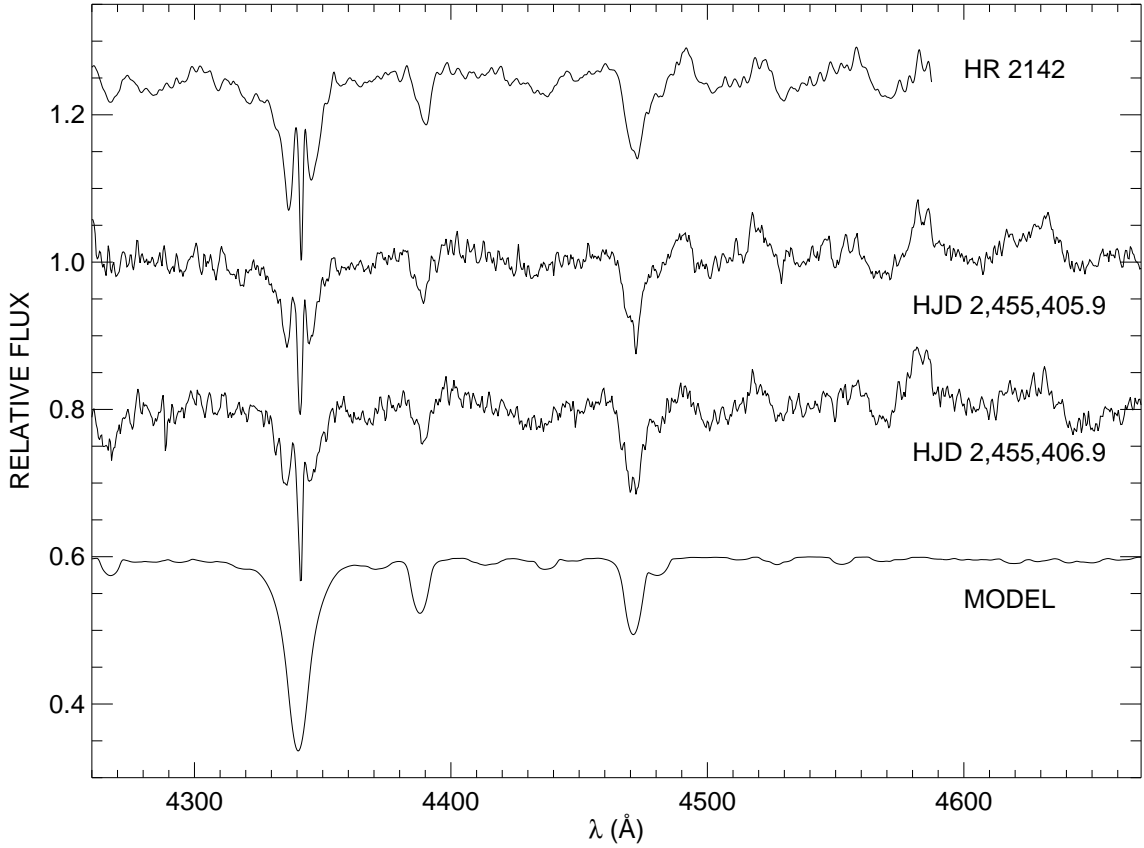}}
\end{center}
\caption{The continuum normalized spectrum of HD~215227 from the first 
and second nights (the latter is offset by $-0.20$ for clarity).  
Above is a similar spectrum of the Be binary HR~2142 (Grundstrom 2007) 
and below is a TLUSTY/SYNSPEC synthetic spectrum for the parameters listed 
in Table~1 (offset by $+0.25$ and $-0.40$, respectively).}
\label{fig1}
\end{figure}

\placefigure{fig1}     

We estimated a number of stellar parameters for the star by comparing 
the observed spectra with B-star model spectra from the 
TLUSTY/SYNSPEC grid of Lanz \& Hubeny (2007).  The best fit model 
parameters from an iterative visual match of the synthetic and 
two observed spectra are listed in Table~1.   
The errors given in Table~1 are based on our judgment of  
the boundary between reasonable and unsuitable solutions, 
and these are intentionally conservative given our incomplete 
understanding of emission contamination in the spectrum.  
We focused first on measurements based upon relative 
spectral intensity since we found clear evidence of 
a systematic line weakening due the presence of disk continuum flux. 
The projected rotational velocity $V\sin i$ was estimated 
from the shape of the \ion{He}{1} $\lambda 4471$ and \ion{Mg}{2} $\lambda 4481$
blend.  The effective temperature $T_{\rm eff}$ 
derivation was based primarily on the \ion{He}{1} $\lambda 4471$ 
to \ion{Mg}{2} $\lambda 4481$ ratio. 
This ratio increases monotonically with increasing $T_{\rm eff}$, 
although the ratio is almost independent of $T_{\rm eff}$ near 
$T_{\rm eff}=21$~kK, which corresponds to the peak of the \ion{He}{1} strength.  
However, we can estimate $T_{\rm eff}$ even in this vicinity by 
the relative strength of the \ion{O}{2} $\lambda 4417$ 
blend, which is stronger at higher $T_{\rm eff}$.  
The gravity $\log g$ was determined by fitting the 
emission-free parts of the Stark broadened, H$\gamma$ line wings.   
We found that the model photospheric spectrum
had line depths that were consistently deeper than the observed ones, 
and this is probably due to disk continuum flux in this spectral region.
Consequently, we renormalized the model spectrum by adding a pure 
continuum component with a disk-to-star, monochromatic flux ratio $F_d/F_\star$. 
This renormalized model spectrum is shown as the lower plot in Figure~1. 
Finally, we estimated the radial velocity $V_r$ for both spectra by 
cross-correlating the observed and model spectra over the wavelength 
range including the emission-free H$\gamma$ wings and \ion{He}{1} $\lambda 4387$.
All these measurements are summarized in Table~1. 

\placetable{tab1}      

The derived stellar parameters suggest a spectral classification of B3~IVne+sh, 
where the temperature -- spectral type relation is taken from B\"{o}hm-Vitense (1981),
the gravity -- luminosity class relation for like stars is adopted from the results of
Huang \& Gies (2008), and ``n'' and ``e'' indicate broad lines and Balmer emission, respectively.
The final suffix ``sh'' is often used to denote the presence of shell lines, 
which is indicated here by the sharp, central absorption in H$\gamma$.   
Note that a similar shell feature may be present in \ion{He}{1} $\lambda 4471$ in 
the spectrum from the first night.  Wolff et al.\ (2007) determined a surprisingly 
low value of projected rotational velocity, $V\sin i = 30$ km~s$^{-1}$, compared 
to our result.  We suspect that their spectrum was obtained at a time when the 
shell lines dominated and that their measurement corresponds to the shell line width. 


\section{Discussion}      

The spectral properties of HD~215227 bear some resemblance to those 
of the gamma-ray binary LS~I~+61~303, which is a Be star in a 26.4960~d orbit
(Aragona et al.\ 2009).  The H$\alpha$ profile in LS~I~+61~303 displays 
systematic variations in the ratio of the violet-to-red ($V/R$) peak emission 
around the time of periastron in this eccentric orbit ($e=0.54$) binary
(Grundstrom et al.\ 2007; McSwain et al.\ 2010).  We find that the 
$V/R$ ratio of the H$\gamma$ emission changed significantly in just one day
(Fig.~1), which is unusually fast for most Be stars (Grundstrom 2007). 
We suggest that such rapid variability might be associated with the 
changing tidal effects of a companion on the disk (especially strong 
near periastron).   

If the disk in HD~215227 is modulated with a binary orbit, then 
the continuum flux from the disk may also show an orbital modulation. 
We found that the disk contributes 
$\approx 33\%$ of the total flux in the optical, so we might expect to find 
some evidence of the orbital modulation in broad-band photometry. 
In fact, Koen \& Eyer (2002) found evidence of a 61~d periodicity in the
{\it Hipparcos} light curve (258 measurements).  There are two other sets of 
photometric measurements from all sky survey experiments.  The first set of 101 
Cousins $I_C$ measurements were made between 2003 and 2007 by The Amateur Sky Survey 
(TASS\footnote{http://sallman.tass-survey.org/servlet/markiv/}; \citealt{dro06}).
The second set of 52 points were made from 1999 to 2000 with the Northern Sky 
Variability Survey (NSVS\footnote{http://skydot.lanl.gov/nsvs/nsvs.php}; 
\citealt{woz04}).   Since these three sets were made with different 
broad-band filters, we assumed that color variations are minimal and 
then simply subtracted the mean magnitude of each set to form a 
combined photometric time series.  These residual magnitudes are 
based upon differences from the average values of $<H_p>=8.78$, 
$<I_C>=8.52$, and $<m_{V{\rm ,ROTSE}}>=9.12$ for the {\it Hipparcos},
TASS, and NSVS photometry sets, respectively.  
A discrete Fourier transform period search 
revealed one significant signal with a period of $60.37 \pm 0.04$~d, 
an epoch of maximum brightness at HJD~2,453,243.3 $\pm 1.8$, and a 
semiamplitude of $0.020 \pm 0.002$ mag.  The resulting light curve (Fig.~2)
displays a low-amplitude, quasi-sinusoidal variation that probably  
corresponds to the binary orbital period.  Note that the small radial velocity
changes we observed over a one day interval are consistent with the small 
variations that are expected for an orbital period this long. 
 
\begin{figure}
\begin{center}
{\includegraphics[angle=90,height=12cm]{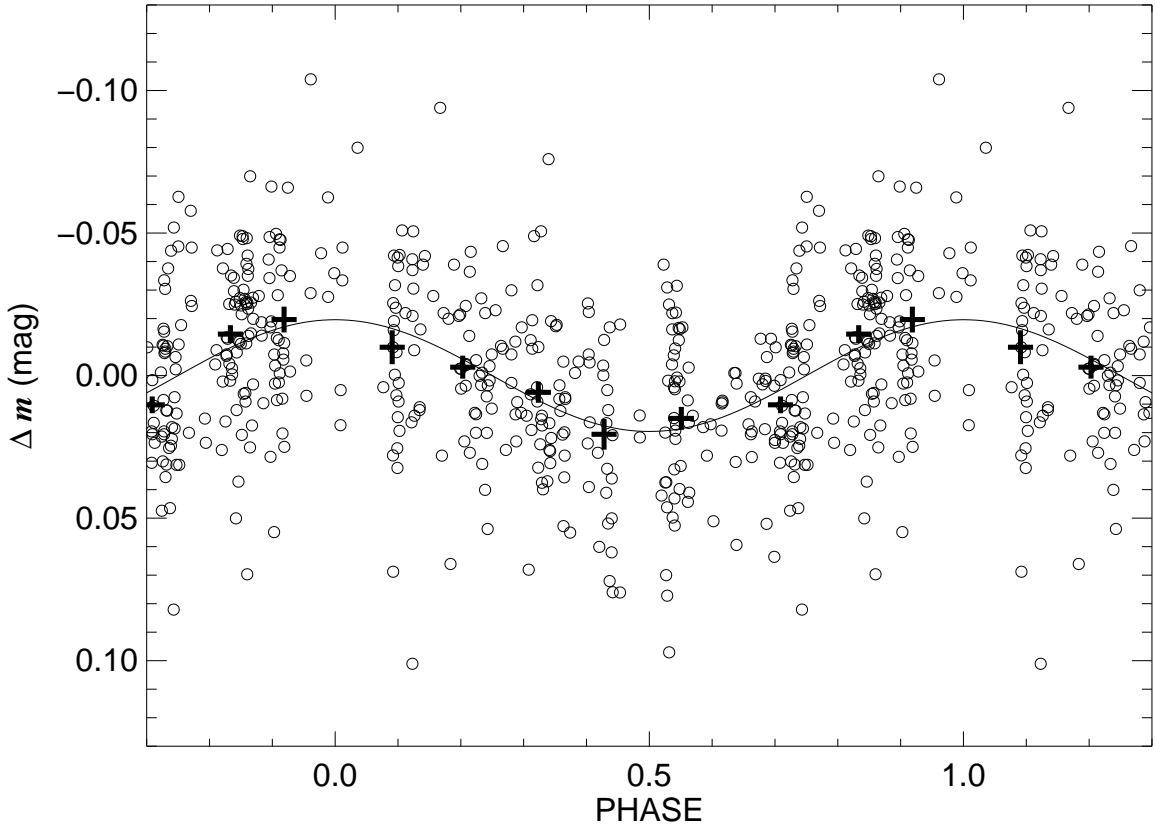}}
\end{center}
\caption{The combined, mean-subtracted, photometry of HD~215227
plotted as a function of cycle phase for a period of 60.37 d.
Dark plus signs show the average $\pm$ one standard deviation of
the mean in each of eight phase bins.}
\label{fig2}
\end{figure}

\placefigure{fig2}     

We note that the Galactic latitude of HD~215227, $b=-12^\circ$, 
suggests that the star is quite far from the Galactic plane, and 
hence it may be a runaway star formed by a supernova (SN) explosion 
in a binary system (Gies \& Bolton 1986; Hoogerwerf et al.\ 2000). 
In this scenario, the SN explosion removes most of the mass of the 
progenitor star, and the stellar companion moves away from the explosion 
site with a speed comparable to its orbital speed at the time of the 
explosion.  If the SN progenitor was the less massive component 
at the time of the explosion and if asymmetries in the SN 
outflow were modest, then the system remains bound as a runaway
star with a degenerate companion.  The runaway velocities may be
large enough to move the system hundreds of parsecs away from 
the location of the SN over the lifetime of the stellar companion. 

The distance to the star is quite uncertain, but we can make an 
estimate by comparing the expected stellar radius from evolutionary
tracks with the angular diameter derived by fitting the spectral 
energy distribution (SED).   We interpolated in the single star 
evolutionary tracks of Schaller et al.\ (1992) to derive mass 
and radius estimates from the derived $T_{\rm eff}$ and $\log g$ 
parameters (Table~1).  We constructed the SED using UV fluxes 
from the TD1 satellite (Thompson et al.\ 1978), Johnson magnitudes
from Mermilliod (1991; transformed to flux using the calibration of
Colina et al.\ 1996), 2MASS magnitudes (Skrutskie et al.\ 2006; 
transformed to flux according to the calibration of Cohen et al.\ 2003), 
and an AKARI 9~$\mu$m measurement (Ishihara et al.\ 2010). 
The SED (Fig.~3) indicates that there is a strong IR-excess from 
the Be star's disk that extends into the optical range. 
In order to fit the photosphere of the star alone, we restricted the 
range to the UV fluxes and $B$-band revised flux without the 
disk contribution (the lower point plotted at 4443 \AA ~in Fig.~3).
We adopted a theoretical flux spectrum from the models of R.\ L.\ 
Kurucz\footnote{http://kurucz.harvard.edu/grids.html}
for solar metallicity, $T_{\rm eff}$ and $\log g$ from Table~1, and 
a microturbulent velocity of 2 km~s$^{-1}$.   
This flux spectrum was fit to the restricted set of observations
with two parameters: the limb-darkened, angular diameter $\theta_{LD}$
and the reddening $E(B-V)$ (assuming a ratio of total-to-selective 
extinction $R_V=3.1$ and the reddening law from Fitzpatrick 1999).
The observed and model emitted fluxes are related by 
$$f_\lambda({\rm obs}) = \theta_{LD}^2 F_\lambda({\rm mod})
 10^{-0.4 A_\lambda}$$
where $A_\lambda$ is the wavelength dependent extinction (Fitzpatrick 1999).  
The first parameter $\theta_{LD}$ acts as a normalizing factor while $E(B-V)$ 
defines how the shape of the SED is altered by the extinction $A_\lambda$. 
The results for these two fitting parameters are given in Table~1, and the
derived $f_\lambda$ spectrum is plotted as a solid line in Figure~3.
We find a very low reddening along this line of sight, $E(B-V)= 0.02$ mag.  
Earlier estimates consistently arrive 
at a higher reddening of $E(B-V)\approx 0.3$ mag (Snow et al.\ 1977;
Neckel \& Klare 1980; Hern\'{a}ndez et al.\ 2005; Zhang et al.\ 2005),
but all of these estimates are based upon the $B-V$ color and ignore the disk 
contribution in the optical that makes the star appear too red (Fig.~3).

Combining the SED-derived angular size $\theta_{LD}$ with 
the evolutionary radius $R_\star$ yields a large distance, $d=2.6 \pm 1.0$~kpc,
which places the star far beyond the nearby Lac~OB1 association (Harris 1955). 
At this distance, HD~215227 resides well below the Galactic plane 
at $z = -0.56 \pm 0.20$~kpc.  This is an extreme distance for a normal OB 
star.  For example, in the Be star kinematical survey by 
Berger \& Gies (2001), the mean distance is $<|z|>=69$~pc and only 
one other star (HD~20340 at $z = -0.71$~kpc) out of a sample of 344
has a distance from the plane as large as that of HD~215227. 
This suggests that HD~215227 is a runaway star that probably obtained its
initial high velocity and current position by a supernova explosion in a binary. 
We adopted the proper motions from van Leeuwen (2007) and the average radial velocity
and distance from Table~1 to estimate the star's current peculiar values 
of tangential $V_{T{\rm p}}$, radial $V_{R{\rm p}}$, and space velocity $V_{S{\rm p}}$
using the method described by Berger \& Gies (2001).  These peculiar 
velocities are measured relative to the star's local standard of rest
by accounting for the Sun's motion in the Galaxy and differential Galactic rotation. 
Our estimates of the peculiar velocities (Table 1)
are not unusually large, but the errors are significant and the star 
may have decelerated in the gravitational potential of the Galaxy. 

\begin{figure}
\begin{center}
{\includegraphics[angle=90,height=12cm]{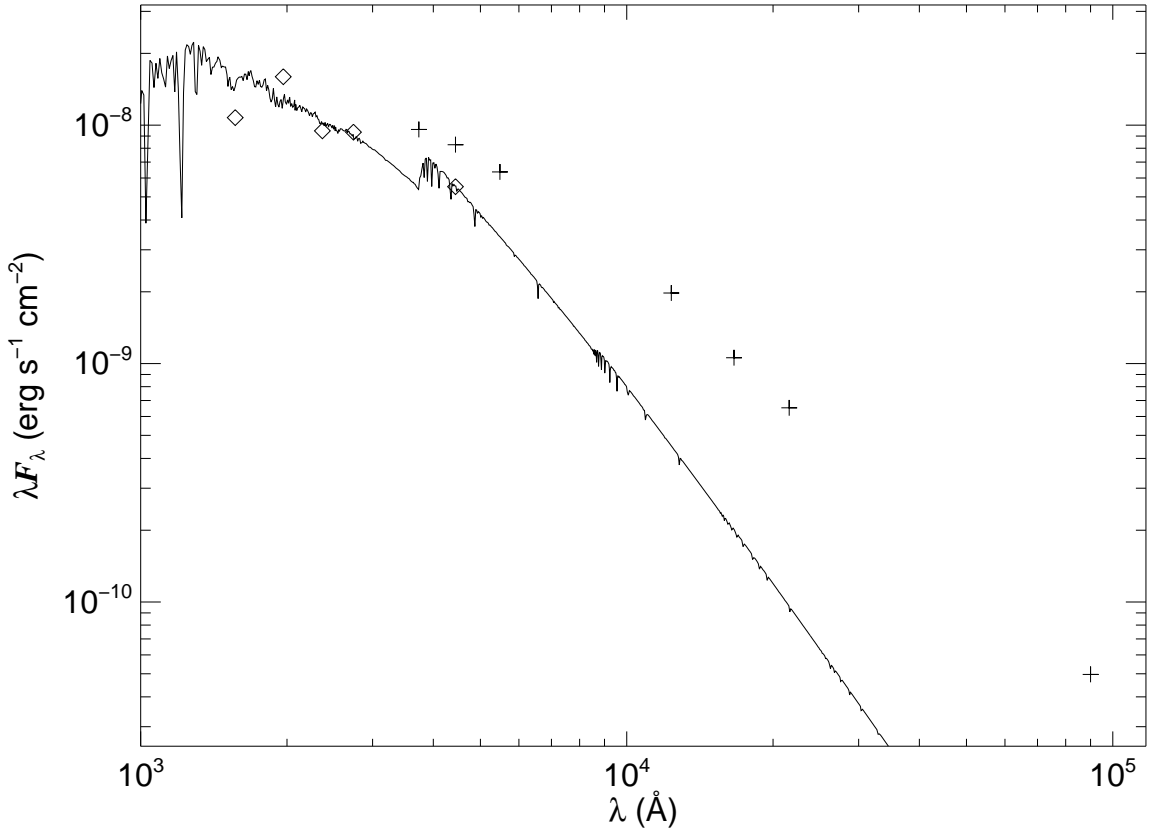}}
\end{center}
\caption{The spectral energy distribution of HD~215227. 
The diamonds indicate the observed fluxes used in the fit
(UV fluxes and the $B$-band corrected flux) while the 
plus signs indicate those fluxes with a significant disk 
contribution that were omitted from the fit.  The solid line 
shows the stellar model flux spectrum for the parameters given in Table 1.}
\label{fig3}
\end{figure}

\placefigure{fig3}     

The similarity of this Be binary to other gamma-ray binaries, 
its probable runaway status, and its close proximity to the gamma-ray 
source location all indicate that HD~215227 may be the optical 
counterpart of AGL~J2241+4454.  We encourage additional spectroscopic
observations to search for orbital motion and to document the emission 
line variability around the orbit.  New X-ray and radio observations 
with good angular resolution will be essential to secure the identification 
of the target (see the case of HESS~J0632+057; Hinton et al.\ 2009; 
Skilton et al.\ 2009).  Furthermore, a search for periodic gamma-ray 
brightening on the 60~d cycle would offer conclusive evidence of the 
connection with the Be star.  If the orbital eccentricity is high, 
the gamma-ray emission may be restricted to a limited part  
of the orbit. 


\acknowledgments

We thank Dr.\ Dmitry Monin and the other staff of the 
Dominion Astrophysical Observatory, Herzberg Institute of Astrophysics, 
National Research Council of Canada, for their assistance
in making these observations possible. 
This material is based on work supported by the
National Science Foundation under Grant AST-0606861.
Institutional support has been provided from the GSU College
of Arts and Sciences and from the Research Program Enhancement
fund of the Board of Regents of the University System of Georgia,
administered through the GSU Office of the Vice President for Research.
We gratefully acknowledge all this support.

{\it Facility:} \facility{DAO:1.85m}




\begin{deluxetable}{lc}
\tablewidth{0pt}
\tablenum{1}
\tablecaption{Stellar Properties\label{tab1}}
\tablehead{
\colhead{Parameter}  &
\colhead{Value}      }
\startdata
$T_{\rm eff}$ (kK)               \dotfill & $19 \pm 3$ \\
$\log g$ (cm s$^{-2}$)           \dotfill & $3.7 \pm 0.2$  \\
$V\sin i$ (km s$^{-1}$)          \dotfill & $300 \pm 50$  \\
$V_r$(HJD 2,455,405.9461) (km s$^{-1}$) \dotfill & $0.2 \pm 1.9$  \\
$V_r$(HJD 2,455,406.9124) (km s$^{-1}$) \dotfill & $2.0 \pm 2.4$  \\
$F_d/F_\star$                    \dotfill & $0.5 \pm 0.3$  \\
$E(B-V)$ (mag)                   \dotfill & $0.02 \pm 0.06$  \\
$\theta_{LD}$ ($10^{-6}$ arcsec) \dotfill & $24 \pm 5$  \\
$M_\star$ ($M_\odot$)            \dotfill & $7.8 \pm 2.0$  \\
$R_\star$ ($R_\odot$)            \dotfill & $6.6 \pm 1.9$  \\
$d$ (kpc)                        \dotfill & $2.6 \pm 1.0$  \\
$z$ (kpc)                        \dotfill & $-0.56 \pm 0.20$  \\
$V_{T{\rm p}}$ (km s$^{-1}$)     \dotfill & $19 \pm 17$  \\
$V_{R{\rm p}}$ (km s$^{-1}$)     \dotfill & $21 \pm 17$  \\
$V_{S{\rm p}}$ (km s$^{-1}$)     \dotfill & $28 \pm 24$  \\
\enddata
\end{deluxetable}



\clearpage


\end{document}